\documentclass[times, twocolumn]{aastex63}
\usepackage{CJK}
\usepackage{graphicx}
\usepackage{enumerate}
\usepackage{amssymb, amsmath}
\usepackage{natbib}
\usepackage{color}
\usepackage{ulem}
\usepackage{dblfnote}
\usepackage{appendix}
\usepackage{hyperref}
\usepackage[stable]{footmisc}
\usepackage{comment}
\usepackage{footnote}
\newcommand{\uat}[2]{\href{http://vocabs.ands.org.au/repository/api/lda/aas/the-unified-astronomy-thesaurus/current/resource.html?uri=http://astrothesaurus.org/uat/#1}{#2 (#1)}}
\usepackage[nodayofweek]{datetime}
\newdateformat{monthyearday}{%
  \THEYEAR\ \monthname[\THEMONTH] \twodigit{\THEDAY}}
  
\accepted{\monthyearday\today}
\submitjournal{AAS Journals}
\shorttitle{Keck/NIRC2 SR 21}
\shortauthors{Uyama et al.}

\begin{document}
\begin{CJK*}{UTF8}{gbsn}
\title{Early High-contrast Imaging Results with Keck/NIRC2-PWFS: The SR~21 Disk}

\author[0000-0002-6879-3030]{Taichi Uyama}
    \affiliation{Infrared Processing and Analysis Center, California Institute of Technology, 1200 E. California Blvd., Pasadena, CA 91125, USA}
    \affiliation{NASA Exoplanet Science Institute, Pasadena, CA 91125, USA}
    \affiliation{National Astronomical Observatory of Japan, 2-21-1 Osawa, Mitaka, Tokyo 181-8588, Japan}
\author[0000-0003-1698-9696]{Bin Ren (任彬)}
    \affiliation{Department of Astronomy, California Institute of Technology, 1200 E. California Blvd., Pasadena, CA 91125, USA}
\author[0000-0002-8895-4735]{Dimitri Mawet}
    \affiliation{Department of Astronomy, California Institute of Technology, 1200 E. California Blvd., Pasadena, CA 91125, USA}
    \affiliation{Jet Propulsion Laboratory, California Institute of Technology, 4800 Oak Grove Dr., Pasadena, CA, 91109, USA}
\author[0000-0003-4769-1665]{Garreth Ruane}
    \affiliation{Jet Propulsion Laboratory, California Institute of Technology, 4800 Oak Grove Dr., Pasadena, CA, 91109, USA}
\author{Charlotte Z. Bond}
    \affiliation{Institute for Astronomy, University of Hawai`i at M\={a}noa, Hilo, HI 96720, USA}
    \affiliation{W. M. Keck Observatory, 65-1120 Mamalahoa Hwy, Kamuela, HI 96743, USA}

\author[0000-0002-3053-3575]{Jun Hashimoto}
    \affiliation{Astrobiology Center of NINS, 2-21-1 Osawa, Mitaka, Tokyo 181-8588, Japan}
\author[0000-0003-2232-7664]{Michael C. Liu}
    \affiliation{Institute for Astronomy, University of Hawaii, 2680 Woodlawn Drive, Honolulu, HI 96822, USA}
\author{Takayuki Muto}
    \affiliation{Division of Liberal Arts, Kogakuin University
2665-1, Nakano-cho, Hachioji-chi, Tokyo, 192-0015, Japan}
\author[0000-0003-2233-4821]{Jean-Baptiste Ruffio}
    \affiliation{Department of Astronomy, California Institute of Technology, 1200 E. California Blvd., Pasadena, CA 91125, USA}
\author[0000-0003-0354-0187]{Nicole Wallack}
    \affiliation{Division of Geological \& Planetary Sciences, California Institute of Technology, Pasadena, CA 91125, USA}

\author[0000-0002-1917-9157]{Christoph Baranec}
    \affiliation{Institute for Astronomy, University of Hawai`i at M\={a}noa, Hilo, HI 96720, USA}
\author[0000-0003-2649-2288]{Brendan P. Bowler}
    \affiliation{Department of Astronomy, The University of Texas at Austin, 2515 Speedway Blvd. Stop C1400, Austin, TX 78712, USA}
\author[0000-0002-9173-0740]{Elodie Choquet}
    \affiliation{Aix Marseille Univ, CNRS, CNES, LAM, Marseille, France}
\author[0000-0002-8462-0703]{Mark Chun}
    \affiliation{Institute for Astronomy, University of Hawai`i at M\={a}noa, Hilo, HI 96720, USA}
\author{Jacques-Robert Delorme}
    \affiliation{Department of Astronomy, California Institute of Technology, 1200 E. California Blvd., Pasadena, CA 91125, USA}
\author[0000-0002-2691-2476]{Kevin Fogarty}
    \affiliation{Department of Astronomy, California Institute of Technology, 1200 E. California Blvd., Pasadena, CA 91125, USA}
\author{Olivier Guyon}
    \affiliation{Subaru Telescope, National Astronomical Observatory of Japan, 650 North A`oh\={o}k\={u} Place, Hilo, HI96720, USA}
    \affiliation{Steward Observatory, University of Arizona, Tucson, AZ 85721, USA}
    \affiliation{Astrobiology Center of NINS, 2-21-1 Osawa, Mitaka, Tokyo 181-8588, Japan}
\author[0000-0003-0054-2953]{Rebecca Jensen-Clem}
    \affiliation{Astronomy \& Astrophysics Department, University of California, Santa Cluz, CA 95064, USA}
\author[0000-0001-6126-2467]{Tiffany Meshkat}
    \affiliation{Infrared Processing and Analysis Center, California Institute of Technology, 1200 E. California Blvd., Pasadena, CA 91125, USA}
\author[0000-0001-5172-4859]{Henry Ngo}
    \affiliation{NRC Herzberg Astronomy and Astrophysics, 5071 West Saanich Road, Victoria, British Columbia, Canada}
\author[0000-0003-0774-6502]{Jason J. Wang}
    \altaffiliation{51 Pegasi b Fellow}
    \affiliation{Department of Astronomy, California Institute of Technology, 1200 E. California Blvd., Pasadena, CA 91125, USA}
\author[0000-0002-4361-8885]{Ji Wang}
    \affiliation{Department of Astronomy, The Ohio State University, 100 W 18th Ave, Columbus, OH 43210, USA}
\author{Peter Wizinowich}
    \affiliation{W. M. Keck Observatory, 65-1120 Mamalahoa Hwy, Kamuela, HI 96743, USA}
\author[0000-0001-7591-2731]{Marie Ygouf}
    \affiliation{Jet Propulsion Laboratory, California Institute of Technology, 4800 Oak Grove Dr., Pasadena, CA, 91109, USA}
\author{Benjamin Zuckerman}
    \affiliation{Department of Physics \& Astronomy, 430 Portola Plaza, University of California, Los Angeles, CA 90095, USA}

\begin{abstract}

High-contrast imaging of exoplanets and protoplanetary disks depends on wavefront sensing and correction made by adaptive optics instruments. Classically, wavefront sensing has been conducted at optical wavelengths, which made high-contrast imaging of red targets such as M-type stars or extincted T Tauri stars challenging. 
Keck/NIRC2 has combined near-infrared (NIR) detector technology with the pyramid wavefront sensor (PWFS). With this new module we observed SR~21, a young star that is brighter at NIR wavelengths than at optical wavelengths. 
Compared with the archival data of SR~21 taken with the optical wavefront sensing we achieved $\sim$20\% better Strehl ratio in similar natural seeing conditions. 
Further post-processing utilizing angular differential imaging and reference-star differential imaging confirmed the spiral feature reported by the VLT/SPHERE polarimetric observation, which is the first detection of the SR~21 spiral in total intensity at $L^\prime$ band.
We also compared the contrast limit of our result ($10^{-4}$ at $0\farcs4$ and $2\times10^{-5}$ at $1\farcs0$) with the archival data that were taken with optical wavefront sensing and confirmed the improvement, particularly at $\leq0\farcs5$.
Our observation demonstrates that the NIR PWFS improves AO performance and will provide more opportunities for red targets in the future.

\end{abstract}

\section{Introduction} \label{sec: Introduction} 

High-contrast imaging has opened a new way of exploring and characterizing exoplanets around young stars \citep[e.g. HR8799 bcde;][]{Marois2008,Marois2010}. This is achieved with large ground-based telescopes by correcting wavefronts of the incident light that are distorted by turbulence of the Earth's atmosphere.
Adaptive optics \citep[AO:][]{Beckers1993} provides real-time corrections to the wavefront distortions using a guide star, which results in a point spread function (PSF) close to a diffraction-limited pattern.
Wavefront sensing has classically been performed at optical wavelengths, in which case the visible brightness of the guide star is very important for the wavefront sensing and correction.
However, for some specific science targets such as M-type stars or extincted young stellar objects (YSOs) that are faint at optical wavelengths while bright in the near-infrared (NIR), wavefront sensing at NIR wavelengths provides better AO performance.
Recently, a NIR Pyramid wavefront sensor (PWFS) was installed in the AO system at Keck-II \citep{Bond2018, Bond2020} as part of the Keck Planet Imager and Characterizer \citep[KPIC:][]{Mawet2016}. Science operations using the NIR-PWFS with Keck/NIRC2 began in early 2019.

EM*~SR~21 (hereafter SR~21) is a YSO in the Ophiuchus star forming region with a distance of $138.4\pm1.1$~pc \citep{gaiadr2}. \cite{herczeg14} found a spectral type of F7 and an age of 10 Myr assuming 121 pc for its distance. \cite{Sallum2019} inferred a spectral type of G3 for assuming the Gaia-based distance.
Its SED shows strong IR excess but lacks emission at $\sim5\micron$, which suggests a transitional disk with an inner gap \citep{Brown2007}.
High-angular resolution studies have shown a large cavity \citep[e.g.][]{vandermarel16}, which together with complicated features such as gaps and spirals suggests possible planet formation within the disk \citep[][]{Sallum2019,muroarena20}.
This YSO is faint at optical wavelengths but bright at NIR wavelengths \citep[e.g., $V$=14.1, $H$=7.5;][]{Cutri2003-2mass, Redbull2018}, making it a suitable testbed for high-contrast imaging with the PWFS to better explore its surrounding disk features and/or companions.
In this study we present the latest observation of SR~21, which is among the earliest science targets high-contrast imaging with with the NIR-PWFS following \cite{wang2020}, who observed PDS~70 using Keck/NIRC2 with the NIR-PWFS. 
Section \ref{sec: Observations and Results} describes our observations, data reduction, and the results. 
Data analysis and discussion including the comparison with the archived Keck/NIRC2 data are described in Section \ref{sec: Discussion}. Finally our observations and results are summarized in Section \ref{sec: Summary}.

\section{Observations and Results} \label{sec: Observations and Results}
\subsection{Observations} \label{sec: Observations}
\begin{figure}
    \centering
    \includegraphics[width=0.45\textwidth]{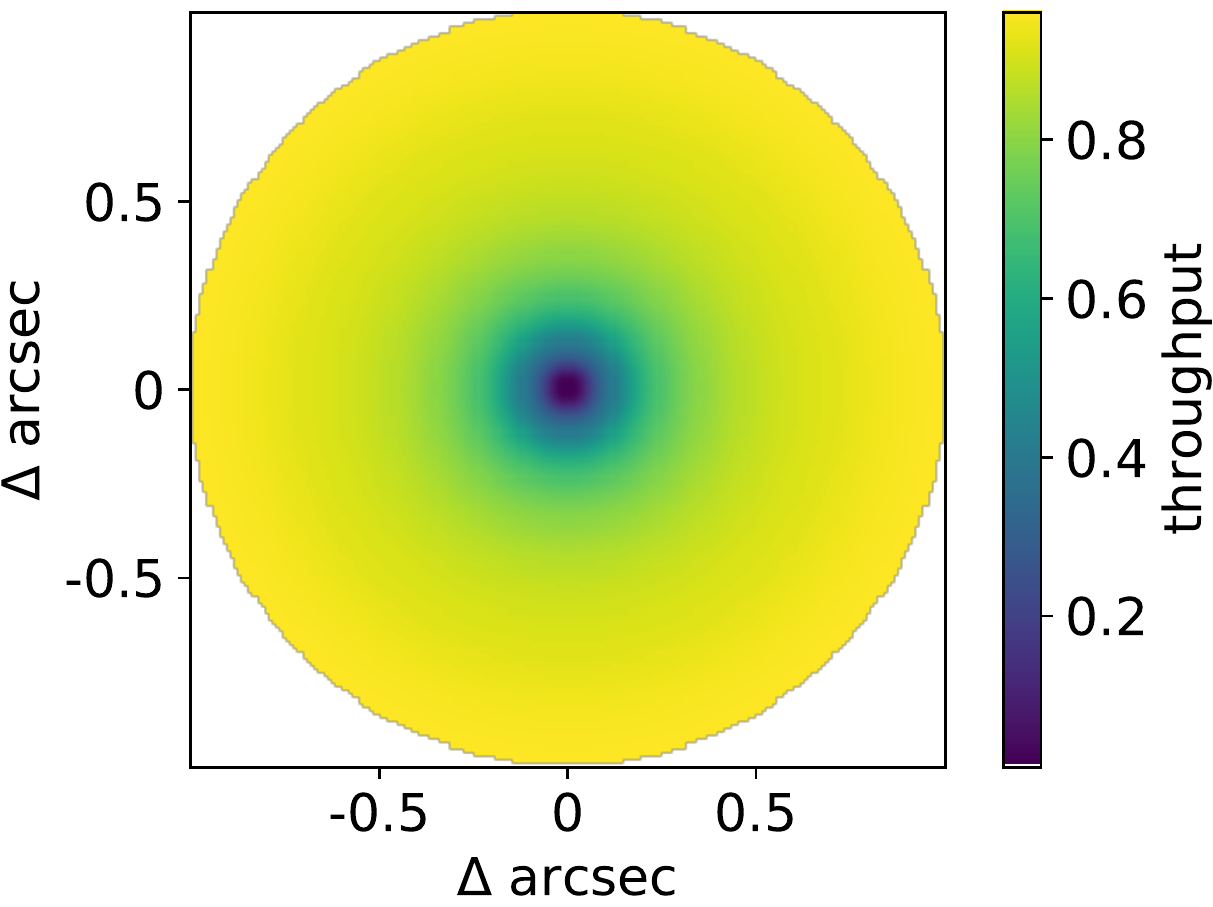}
    \caption{Simulated throughput map around the vortex mask. The off-axis SR~21 data used in this study were taken with an offset of $\sim0\farcs28$ from the vortex mask, which affected the throughput of the off-axis PSF.}
    \label{fig: throughput map}
\end{figure}

We observed SR 21 on 2020 May 31 UT with Keck/NIRC2+PWFS in $L'$ band combined with the vector vortex coronagraph mask \citep[inner working angle: IWA$\sim$100 mas;][]{serabyn2017,mawet2017} and angular differential imaging \citep[ADI:][]{marois06}.
The wavefront sensing was carried out at $H$ band.
The total exposure time is $4110$~sec (0.3 sec per exposure $\times$ 100 coadds $\times$ 137 frames) and the parallactic angle change is $56\fdg31$.
We also took an off-axis image offset by $\sim0\farcs28$ from the vortex mask, which is comprised of 100 coadds each having an integration time of 0.0075 sec.
Typical full width at half maximum (FWHM) was measured to be 8.2 pix (or 81.8 mas with a pixel scale of 9.972 mas/pix). We compared this unsaturated PSF with the simulated ideal PSF at $L^{\prime}$ band, where we took into account throughput loss by the vortex mask (see Figure \ref{fig: throughput map}), and we obtained a Strehl ratio of $\sim87\%$ (approximately equivalent to the root mean square (RMS) of the wavefront errors $\simeq$200 nm).
The current throughput of the NIRC2-PWFS is $\sim9\%$ and the maximum elevation of SR~21 was $\sim45^\circ$ during our observations, which affected the AO performance compared with the ideal case \citep[][]{Bond2018, Bond2020}.
The data were taken as a part of a shared-risk program using the NIR-PWFS (PI: Dimitri Mawet) which was aimed at science verification with redder targets.
In this program we did not observe suitable reference stars for reference-star differential imaging \citep[RDI:][]{ruane2019} and thus we decided to use other data sets (see Table \ref{tab: obs log} and Section \ref{sec: RDI}) that have the most comparable luminosity at $H$ and $W1$ bands to SR~21 among targets observed at this epoch as reference PSFs for the RDI reduction.

\begin{table*}
    \centering
    \caption{Observing Log}
    \begin{tabular}{cccccc}
         target & date [UT] & total exposure [s] & seeing [$^{\prime\prime}$]$^1$ & $H$ [mag]$^2$ & $W1$ [mag]$^3$  \\ \hline\hline
         SR 21 & 2020 May 31 & 4110 & 0.50 & 7.5 & 6.1 \\
         WaOph 6 & 2020 May 30 & 1230 & 0.49 & 7.6 & 7.6 \\
         WSB 52 & 2020 Jun 2 & 5040 & 0.47 & 9.2 & 7.5  \\
    \end{tabular}
    \label{tab: obs log}
\tablecomments{
1. Mean DIMM seeing at the summit of Mauna Kea on each date. We note that during these observations part of the seeing information was not recorded. It is possible that the NIRC2 data were taken at different seeing conditions.\\
2. 2MASS photometry \citep{Cutri2003-2mass}. \\
3. AllWISE photometry \citep{Cutri2013allwise}.}
\end{table*}

\subsection{Data Reduction and Results} \label{sec: Data Reduction and Results}
We first carried out preprocessing, including bad pixel correction, flat fielding, sky subtraction and image registration following \citet{xuan18} and \citet{ruane2019}. We then performed post-processing to subtract the stellar halo.
We investigate two approaches of post-processing here: RDI to investigate the faint disk features and inner potential companions, and ADI to investigate outer potential companions. 
For both approaches, we used the {\tt pyKLIP} package\footnote{\url{https://pyklip.readthedocs.io/en/latest/index.htmll}} \citep{pyklip} that adopted Karhunen-Lo\`eve Image Projection \citep[KLIP:][]{soummer12, amara12} to produce the most likely reference PSF for the target exposure.

\subsubsection{RDI} \label{sec: RDI}
As we did not take suitable data sets for the reference PSF, the RDI results can be affected by the quality of other stars in the reference library.
In addition to the SR~21 data, we included WaOph 6 and WSB 52 data sets (PI: Charlotte Bond, Dimitri Mawet) in the PSF library (see Table \ref{tab: obs log} for the observing logs).
We decided to use these two stars as reference stars because their $W1$ magnitudes are the closest to SR~21 among all targets taken on and around 2020 May 31.
These two stars are young stellar objects (YSOs) surrounded by protoplanetary disks, which have been resolved by \citet{huang2018-dshap2, huang2018-dsharp3}. The NIRC2 $L'$-band PSF for these targets may be influenced by the disks, which could impact the RDI reduction at particularly inner separations.

As mentioned above, the reference stars (WaOph 6 and WSB 52) used in this study are not ideal for the RDI reduction of SR~21 because 1) their $W1$ magnitudes are not exactly the same, 2) observational conditions may vary from target to target, and 3) they possibly include the protoplanetary disk signals at $L^{\prime}$ band, which can degrade the efficiency and output of the post-processing with RDI.
We first conducted frame selection among the PSF library by taking advantage of the mean square error \citep[MSE:][]{wang2004-mse,ruane2019}.
We evaluated MSE scores at separations $\leq$ 390 mas ($\sim$5 $\lambda$/D) and removed the worst 20\% of frames in each data set.
After the frame selection the field of view (FoV) was cropped into $2\farcs0\times2\farcs0$, which is used to make the reference PSF. We performed RDI reduction including all the science data sets and the selected good reference data sets using {\tt pyKLIP}.

\begin{figure*}
\begin{tabular}{ll}
\begin{minipage}{0.45\hsize}
    \centering
    \includegraphics[width=0.9\textwidth]{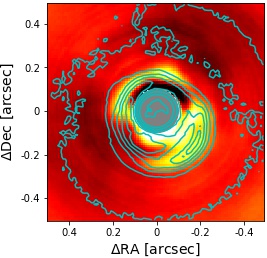}
\end{minipage}
\begin{minipage}{0.45\hsize}
    \centering
    \includegraphics[width=0.9\textwidth]{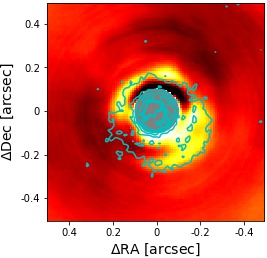}
\end{minipage}
\end{tabular}
\caption{RDI-reduced image of SR 21 (KL=5) overlaid with SPHERE-PDI contours (cyan, left: $H$-band with coronagraph, right: $J$-band without coronagraph). Color scale is arbitrarily set to clearly show the spiral feature. The gray-masked area indicates the NIRC2 vortex coronagraph. North is up and East is left. The dark regions are due to over-subtraction by the RDI reduction, which is probably impacted by the imperfect background subtraction and/or the non-ideal reference stars.}
\label{fig: RDI}
\end{figure*}

Figure \ref{fig: RDI} shows the RDI-reduced result (Karhunen-Lo\`eve - the number of basis vector; KL=5) in total intensity, which is overlaid with contours from SPHERE/IRDIS $H$-band coronagraphic (left) and $J$-band non-coronagraphic (right) observations. We reduce the public SPHERE observations of SR~21 on UT 2018 March 1 (ESO GTO program 1100.C-0481(Q), PI: J.-L.~Beuzit) using {\tt IRDAP} \citep{vanholstein17, vanholstein20}. 
The bright extended feature at position angle (PA) between $\sim170^\circ$ and $\sim270^\circ$ is clearly co-located with the spiral detected by SPHERE \citep['Spiral 1' in][]{muroarena20} and this detection is one of several cases that the spiral feature is clearly resolved in $L'$-band \citep[e.g. HD 142527, HD 100546, MWC 758, and CQ Tau;][]{rameau2012,currie2015,Reggiani2018,uyama2020}.
To test the potential contamination of the disk feature around the reference stars in $L^{\prime}$ band, we checked the archival VLT/SPHERE $H$-band polarimetric data set of WaOph6 taken on UT 2018 June 22 (ESO GTO program 1100.C-0481(Q), PI: J.-L.~Beuzit). The SPHERE data set presents very faint feature of the disk surface of WaOpH~6, and the SPHERE-based SR~21 spiral arm is $\sim$30 times brighter than the WaOph~6 disk at the same position, indicating that the contamination of the WaOph~6 disk on the NIRC2 PSF of SR~21 should be relatively very small. 
For WSB~52 there is no public SPHERE data in the VLT archive.

However, our observation did not confirm other features the SPHERE observations reported. The inner area of the processed image has regions with both positive and negative intensity. Part of the positive areas might correspond to the real disk features but they are likely affected by the poor RDI reduction due to the non-ideal reference stars.

\subsection{ADI} \label{sec: ADI}
We conducted basic ADI reduction using {\tt pyKLIP} using all of the 137 frames. We divided the NIRC2 FoV into 9 annuli $\times$ 4 subsections to produce the most likely reference PSF at each area. We also assumed the real astrophysical signal, if any, moves 1 pix from frame to frame to avoid heavy self-subtraction caused by the ADI reduction.

\begin{figure*}
\begin{tabular}{ll}
\begin{minipage}{0.45\hsize}
    \centering
    \includegraphics[width=0.9\textwidth]{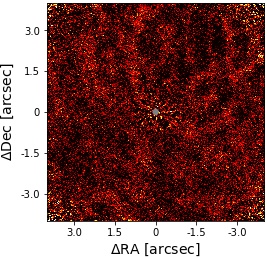}
\end{minipage}
\begin{minipage}{0.45\hsize}
    \centering
    \includegraphics[width=0.9\textwidth]{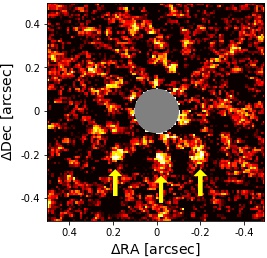}
\end{minipage}
\end{tabular}
\caption{ADI-reduced image of SR 21 (KL=20, left: larger FoV, right: zoomed-in image). The marginal point sources are indicated by yellow arrows (see also Section \ref{sec: ADI}). The Southwest and South (right and middle) sources coincide with the spiral feature. The gray-masked area indicates the NIRC2 vortex coronagraph. North is up and East is left.}
\label{fig: ADI}
\end{figure*}

\begin{figure*}
    \centering
    \includegraphics[width=0.9\textwidth]{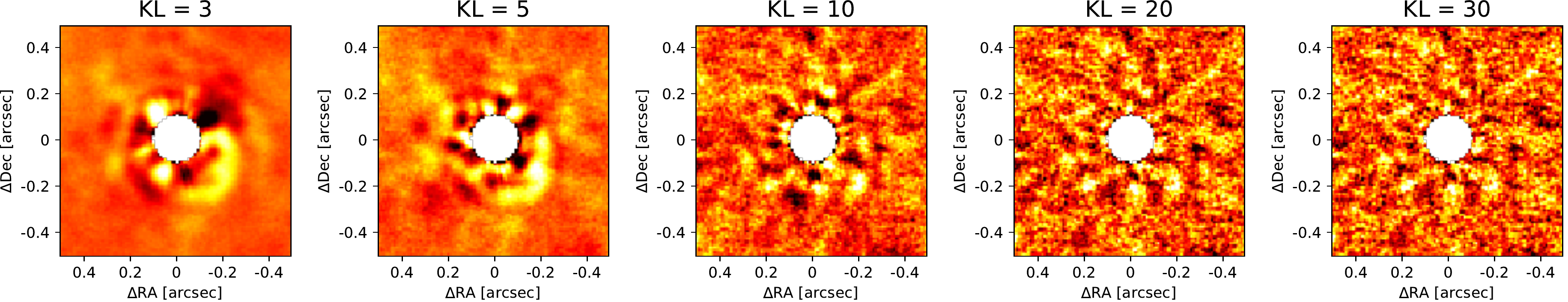}
    \caption{Comparison of the ADI-result at different KLs. The small-KL image (KL$<10$) is more sensitive to an extended feature such as the spiral while the large KL is more sensitive to a point source.}
    \label{fig: ADI KLs}
\end{figure*}

\begin{figure*}
\begin{tabular}{ll}
\begin{minipage}{0.45\hsize}
    \centering
    \includegraphics[width=0.9\textwidth]{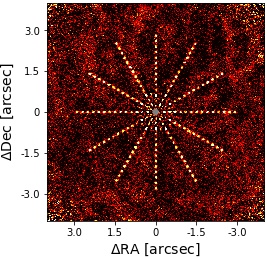}
\end{minipage}
\begin{minipage}{0.45\hsize}
    \centering
    \includegraphics[width=0.9\textwidth]{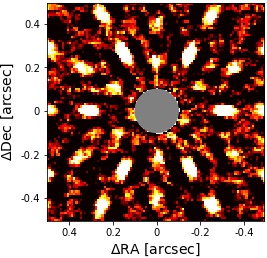}
\end{minipage}
\end{tabular}
\caption{Same as Figure \ref{fig: ADI} with injected fake sources at separations ranging $0\farcs1$-$3\farcs0$. The radially elongated shapes of the fake sources are affected by self-subtraction in the ADI reduction along the azimuthal direction.}
\label{fig: ADI with fake PSFs}
\end{figure*}

Figure \ref{fig: ADI} shows the ADI result (KL=20) for the larger FoV (left) and for the zoomed-in image (right) to compare with the RDI result. We did not detect any significant sources within $\sim3^"$. 
With a smaller number of KL we also see the spiral feature in the ADI-reduced image (Figure \ref{fig: ADI KLs}).
In the zoomed-in image of Figure \ref{fig: ADI}, there are two marginal point-like sources to the Southwest and South that are coincident with the spiral feature and stable among KLs between 10 and 30.
However, their signal to noise ratios (SNRs) are not large enough to robustly claim a detection ($\sim$3.5, 3.7, respectively). 
There is another marginal source outside the spiral feature at Southeast but this is fainter than the sources mentioned above.
In this study, we regard them as the residuals of the spiral feature or uncorrected speckle noise and future observation that can explore deeper. contrast at $\sim0\farcs2-0\farcs3$ will help to identify the nature of these sources especially in the context of planet-disk interactions in the SR~21 system. 
We also injected fake sources at separations ranging $0\farcs1$-$3\farcs0$ and at a variety of position angles to test whether the inject sources are reliably reproduced after the ADI reduction. We altered the contrast levels of the fake sources as a function of radius ($7.5\times10^{-4}$ at $r\leq0\farcs3$, $1.5\times10^{-4}$ at $0\farcs3<r\leq0\farcs7$, and $5.0\times10^{-5}$ at $r>0\farcs7$) because a bright source can result in large flux loss. 
Figure \ref{fig: ADI with fake PSFs} shows a result of the injection test. 
Although part of the disk features may affect the outcome of the ADI reduction, the injection test shows that the point sources can be reliably recovered.

\section{Discussion} \label{sec: Discussion}
\subsection{Spiral Feature} \label{sec: Spiral Feature}
The disk feature detected in $L^{\prime}$ band may be explained by two scenarios: scattering or self-luminosity. 
If the spiral feature is self-luminous it may be a result of heating a shock around the protoplanet, which would indicate planet-disk interactions \citep[e.g.][]{Lyra2016}. 

We use our RDI result to extract the surface brightness of the spiral feature because RDI causes only modest signal loss \citep[e.g. $\lesssim15\%$ with the PCA-based RDI reduction of the disk feature;][]{Uyama2020-HD34700} compared with self-subtraction of ADI, particularly at such small separations.
We converted the pixel count into surface brightness by comparing the photometric result of the off-axis PSF (aperture radius=3.5$\times$FWHM) with the {\it Spitzer}/IRAC-3.6$\mu$m value \citep[1.23$\pm$0.01 Jy;][]{Gutermuth2009-IRAC}.
We traced the spiral feature and investigated several areas, which measures $\sim$60, 100, 80, 70 mJy~arcsec$^{-2}$ at PA=180$^\circ$, 210$^\circ$, 240$^\circ$, and 270$^\circ$, respectively\footnote{We do not present the error bars for these values because we are not able to evaluate systematic errors caused by the non-ideal reference stars. The background noise estimated from standard deviation in an annular area at $\sim1\farcs0$ in the RDI-reduced image corresponds to 6 mJy/arcsec$^2$.}.
Assuming these values reflect thermal emissions from the optically-thick surface of the spiral, the typical temperature of the spiral corresponds to $\sim200$ K. On the other hand, assuming $L_{\rm SR 21}=11 L_\odot$ \citep{Francis2020} and $0\farcs23$ for the typical separation of the spiral, the effective temperature at the spiral is estimated at $\lesssim80$ K and the derived temperature from our RDI result suggests the scattering effect or another heating mechanism \cite[e.g. the shock scenario;][]{Lyra2016}.

We also calibrated the SPHERE $H$-band ${\rm Q_\phi}$ image following the {\tt IRDAP} pipeline. We measured that polarimetric intensity at the spiral surface is $35$ -- $40$ mJy arcsec$^{-2}$ along the spine, with an error of $\pm1.3$~mJy arcsec$^{-2}$, which is defined as standard deviation in an annular area (at radii between 175 and 225 mas) of the calibrated SPHERE ${\rm U_\phi}$ image that is smoothed with a Gaussian kernel ($\sigma=2$ pix). 
By comparing these values with radiative transfer simulations one can infer the scattering characteristics and the possibility of thermal emission at the spiral surface, but we note that simply comparing our $L^\prime$-band total intensity with the SPHERE $H$-band polarimetric imaging result leaves uncertainty on polarization ratio. 
Obtaining total intensity and polarization intensity at the same wavelength is useful to characterize the scattering mechanisms at the disk surface in detail.

As mentioned in Section \ref{sec: Observations and Results}, the reference stars we used in this study are not suitable for SR~21 and thus the extracted surface brightness may be biased.
Future follow-up observations with more suitable reference stars will help to better investigate the whole disk feature including the spiral in $L^\prime$ band.

\subsection{Detection Limits} \label{sec: Detection Limits}

\begin{figure*}
    \begin{minipage}{0.5\hsize}
    \centering
    \includegraphics[width=0.9\textwidth]{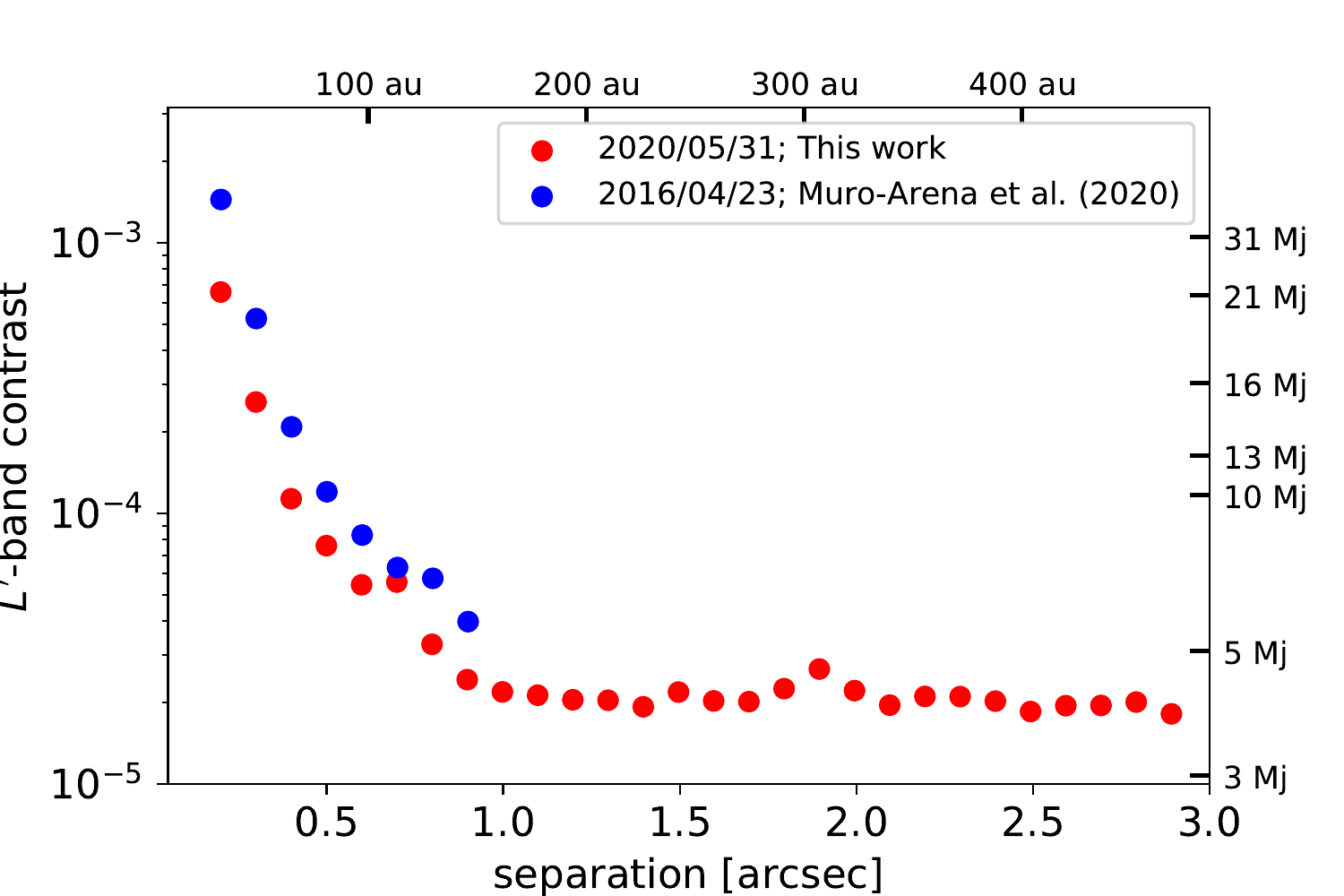}
    \caption{5$\sigma$ contrast limits of the ADI (KL=20) result of our data (red) compared with approximate values of the archival data (blue). We adopted $L^{\prime}$=5.8 mag, which is different from \cite{muroarena20}, for the SR~21 luminosity. For the plot of the previous data we checked the contrast limit presented in Figure A.1 of \cite{muroarena20} and selected several points between $0\farcs2-0\farcs9$.
    }
    \label{fig: contrast}
    \end{minipage}
    \begin{minipage}{0.5\hsize}
    \centering
    \includegraphics[width=0.85\textwidth]{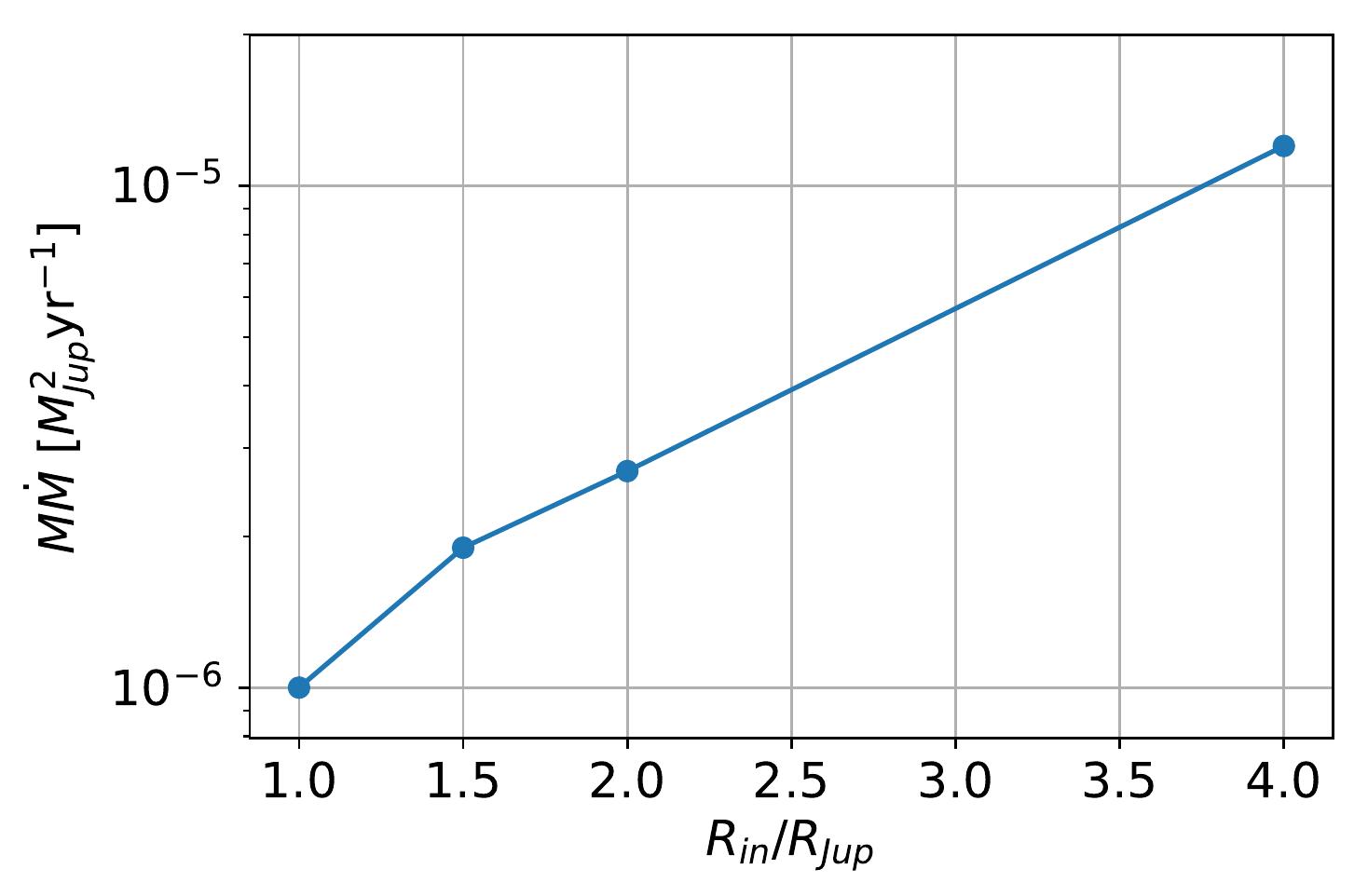}
    \caption{Comparison of our detection limits with the circumplanetary-disk model \citep{Zhu2015} at 44~au ($\sim0\farcs3$) where \cite{muroarena20} predicted a $\lesssim1\ M_{\rm Jup}$ planet. Assuming $\sim1\ M_{\rm Jup}$ for the mass the vertical axis (the product of mass $M$ and mass accretion rate $\dot{M}$) is interpreted as the accretion rate and thus this comparison can be used to constrain the accretion rate. The horizontal axis corresponds to inner radius of the circumplanetary disk ($R_{\rm in}$).
}
    \label{fig: detection limit CPD}
    \end{minipage}
\end{figure*}

We convolved the ADI-reduced image by a circular aperture with a radius of FWHM/2 and then calculated the radial noise profile, which is then compared with the unsaturated PSF of SR 21. 
We calibrated this radial profile by the flux loss ratio at each separation estimated from the injection test result.
Figure \ref{fig: contrast} shows our 5$\sigma$ contrast limit: we achieved 10$^{-4}$ at $0\farcs4$ and $2\times10^{-5}$ at $1\farcs0$. With the COND03 model \citep{baraffe2003} assuming 10~Myr overlaid in Figure \ref{fig: contrast}, we set a constraint on $\sim16\ M_{\rm Jup}$, $10\ M_{\rm Jup}$, and $5\ M_{\rm Jup}$ at $\sim40$~au, 55~au, and 110~au, respectively.
We assumed 5.8 mag at $L^\prime$ band, which is different from \cite{muroarena20}. Their assumption of 6.8 mag is equivalent to 0.48 Jy at 3.8$\micron$ and this value is fainter than photometric values at adjacent wavelengths \citep[e.g. $\sim1.1-1.3$ Jy at 3.5-4.6$\micron$;][]{Gutermuth2009-IRAC,Cutri2013allwise,Schlafly2019-unWISE}. Our assumption of 5.8 mag is equivalent to the {\it Spitzer}/IRAC-3.6 $\micron$ value.
Our mass limit is consistent with an updated mass limit of the archival data from Figure A.1 of \cite{muroarena20} (in private communications).
We also took into account extinction by the interstellar medium
considering $A_{V}=6.2$ \citep{herczeg14} and wavelength dependence of extinction \citep[$A_{\lambda}\propto\lambda^{-1.75}$:][]{Draine1989}, which corresponds to $A_{L^\prime}=0.2$.

The evolutionary model we adopted in Figure \ref{fig: contrast} does not include the accreting mechanism. Given that the gas in the SR~21 disk has not been fully depleted at the predicted location the potential protoplanet would likely be actively accreting. So we also compare our contrast limits with a 'circumplanetary-disk' model \citep{Zhu2015}, with which the $L^\prime$-band luminosity is calculated as a function of the product of planet mass ($M$) and accretion rate ($M\dot{M}$) and inner radius of the circumplanetary disk ($R_{\rm in}$). 
We follow \cite{Ruane2017} who investigated constraints on the potential accreting planets in the TW Hya disk. 
\cite{muroarena20} predicted a $\lesssim 1M_{\rm Jup}$ protoplanet at 44~au and PA $\sim11^\circ$ and we use our contrast limit to set a constraint on the mass accretion rate onto this potential protoplanet.
Figure \ref{fig: detection limit CPD} shows the comparison between our contrast limit at $0\farcs3$ (contrast: 2.6$\times10^{-4}$) and the cirucmplanetary-disk model at given inner radii ($1R_{\rm Jup}\leq R_{\rm in} \leq 4R_{\rm Jup}$).
Even if we assume 1 $M_{\rm Jup}$ for the potential protoplanet as predicted in \cite{muroarena20}, we are not able to solve the degeneracy between the mass accretion rate and the inner radius of the circumplanetary disk because no observational evidence of the circumplanetary disk in the SR~21 disk has been reported. 
Therefore our detection limit could set an upper limit of the accretion rate to $\dot{M}\lesssim1.2\times10^{-5}\ M_{\rm Jup}$yr$^{-1}$ by referring to a $R_{\rm in}=4R_{\rm Jup}$ case in the circumplanetary-disk model.

These estimations of the mass/accretion-rate limits do not include potential extinction of the SR~21 disk itself. This leaves uncertainty on the mass estimation but this may not be as important because the $L^{\prime}$-band detection limits are less subject to the extinction than shorter NIR wavelengths (e.g. $JHK$ bands).

\subsection{Comparison with the Archival Data} \label{sec: Comparison with the Archival Data} 

Compared with the archival Keck/NIRC2 $L'$-band data taken on 2016 April 23 \citep[PI: Van der Marel, see Appendix A of][]{muroarena20}, we detected the spiral feature in both of the RDI (Figure \ref{fig: RDI}) and ADI (KL=3 and 5 at Figure \ref{fig: ADI KLs}) reductions, while \cite{muroarena20} did not report any disk features in the archival Keck data.
Our detection limit also achieves a better contrast level particularly at inner separations than the archival data (see Figure \ref{fig: contrast} for the comparison of the detection limits).
At separations $\geq0\farcs5$ the difference of the contrast limits is likely related to difference of the exposure time (4110 sec and 2160 sec for our observation and the previous observation, respectively).
We also note that difference of field rotation ($56\fdg31$ and $18\fdg19$, respectively) may also affect the detection limits at inner separations because a ratio of flux loss made by the ADI reduction can be affected by the rotation angle.

The main difference is that we used the vortex coronagraph and NIR-PWFS while the archival data were taken with the optical wavefront sensing and no coronagraph mask. We took into account the effect of the vortex mask when we measured the Strehl ratio (see Section \ref{sec: Observations}).
The observing conditions are also different - typical DIMM seeings were $0\farcs50$ and $0\farcs77$ on 2020 May 31 and 2016 April 23 respectively.
We measured the Strehl ratio of the archival data to be $\sim$69\%, which is smaller than ${\sim}87\%$ that we measured for our observation. 
Although seeing was slightly larger on 2016 April 23 when the archival data were taken, the majority of the additional wavefront errors due to the increased seeing should be outside the AO correction radius.
The increase of the Strehl ratio by $\sim$20\% is likely produced by the NIR-PWFS suggesting that the NIR-PWFS meaningfully improved the AO performance of Keck/NIRC2 on such a red target.

\section{Summary} \label{sec: Summary}

We have presented the latest Keck/NIRC2 $L^\prime$-band observation of SR~21.
The observation was conducted with a NIR ($H$ band) PWFS and the Strehl ratio is measured to be $\sim$87\% (approximately 200 nm for RMS of the wavefront errors). 
We used the {\tt pyKLIP} package to post-process the observation utilizing RDI and ADI techniques.
In the RDI-reduced image, we confirmed the spiral feature in total intensity at $L^\prime$ band, which is consistent with the VLT/SPHERE polarimetric observation.
However, we did not observe suitable reference stars for SR~21 but instead used WaOph 6 and WSB 52, which might also include disk signals. Therefore our RDI result may have been affected by these non-ideal reference stars.
Our ADI reduction did not detect any convincing companion candidates but we confirmed the spiral feature at small KLs (KL $\leq$ 5).
In addition to the spiral feature at small KLs, we see two point-like sources with SNRs of $\sim3.5-3.7$ at KL = 10--30 but they may be residuals of the spiral feature.
We calculated a 5$\sigma$ contrast limit from the ADI result, which achieved $10^{-4}$ at $0\farcs4$ and $2\times10^{-5}$ at $1\farcs0$. Assuming COND03 model and 10 Myr, we converted the contrast limit into a mass limit - 16~$M_{\rm Jup}$, 10~$M_{\rm Jup}$, and 5~$M_{\rm Jup}$ at 40~au, 50~au, and 110~au, respectively.
Compared with the archival data taken with optical wavefront sensing we achieved the better Strehl ratio and better contrast level, which suggests that the NIR PWFS improved the AO performance of Keck/NIRC2 and will provide more opportunities for red targets.
As the spiral feature is affected by the bias of the reduction method and the reference stars used in this study, future observation of this system will provide better RDI results and enable a detailed discussion on the surface brightness distribution of the protoplanetary disk.

\facility{Keck:II (NIRC2)}
\software{{\tt pyKLIP} \citep{pyklip}, {\tt IRDAP} \citep{vanholstein17, vanholstein20}}

\acknowledgments
The authors would like to thank the anonymous referee for the constructive comments to improve the clarity of the manuscript.
We thank Christian Ginski for sharing the updated detection limit of the archival Keck/NIRC2 data.
Some of the data presented herein were obtained at the W. M. Keck Observatory, which is operated as a scientific partnership among the California Institute of Technology, the University of California and the National Aeronautics and Space Administration. The Observatory was made possible by the generous financial support of the W. M. Keck Foundation.
Based on observations collected at the European Organisation for Astronomical Research in the Southern Hemisphere under ESO programme 1100.C-0481(Q).
This research has made use of NASA's Astrophysics Data System Bibliographic Services.
This research has made use of the SIMBAD database \citep{simbad}, operated at CDS, Strasbourg, France. This research has made use of the VizieR catalogue access tool, CDS,  Strasbourg, France (DOI:  \href{https://doi.org/10.26093/cds/vizier}{10.26093/cds/vizier}). The original description of the VizieR service was published in A\&AS 143, 23 \citep{ochsenbein00}.

TU acknowledges JSPS overseas research fellowship. ML acknowledges funding from the National Science Foundation under grants AST-1518339. The Keck infrared pyramid wavefront sensor was developed with support from the National Science Foundation under grants AST-1611623 and AST-1106391, as well as the Heising Simons Foundation under the Keck Planet Imager and Characterizer project. This research is partially supported by NASA ROSES XRP, award 80NSSC19K0294.
Part of this work was carried out at the Jet Propulsion Laboratory, California Institute of Technology, under contract with the National Aeronautics and Space Administration (NASA).

We wish to acknowledge the critical importance of the current and recent Mauna Kea Observatories daycrew, technicians, telescope operators, computer support, and office staff employees, especially during the challenging times presented by the COVID-19 pandemic. Their expertise, ingenuity, and dedication is indispensable to the continued successful operation of these observatories.
The authors wish to acknowledge the very significant cultural role and reverence that the summit of Mauna Kea has always had within the indigenous Hawaiian community. We are most fortunate to have the opportunity to conduct observations from this mountain.

\newpage
\bibliography{library}                                    
\end{CJK*}
\end{document}